\documentclass{jpp}
\usepackage{graphicx}

\usepackage[utf8]{inputenc}
\usepackage[T1]{fontenc}
\usepackage{amsmath}
\usepackage{color} %Only needed for the note commands added.
\newcommand{\gke}{\texttt{Gkeyll}}
\newcommand{\pic}{\texttt{p3d}}

\shorttitle{Noise-Induced Magnetic Field Saturation}
\shortauthor{J. Juno, M. Swisdak, J. M. TenBarge, V. Skoutnev, and A. Hakim}

\title{Noise-Induced Magnetic Field Saturation in Kinetic Simulations}

\author{J. Juno\aff{1}
  \corresp{\email{jjuno@terpmail.umd.edu}},
  M. Swisdak\aff{1},
  J. M. TenBarge\aff{2},
  V. Skoutnev\aff{2},
  \and A. Hakim\aff{3}}

\affiliation{\aff{1}IREAP, University of Maryland, College Park, MD 20742, USA
\aff{2}Department of Astrophysical Sciences, Princeton University, Princeton, NJ 08544, USA
\aff{3}Princeton Plasma Physics Laboratory, Princeton, NJ 08543, USA}

\begin{document}

\maketitle

\begin{abstract}
Monte Carlo methods are often employed to numerically integrate kinetic equations, such as the particle-in-cell method for the plasma kinetic equation, but these methods suffer from the introduction of counting noise to the solution. We report on a cautionary tale of counting noise modifying the nonlinear saturation of kinetic instabilities driven by unstable beams of plasma. We find a saturated magnetic field in under-resolved particle-in-cell simulations due to the sampling error in the current density. The noise-induced magnetic field is anomalous, as the magnetic field damps away in continuum kinetic and increased particle count particle-in-cell simulations. This modification of the saturated state has implications for a broad array of astrophysical phenomena beyond the simple plasma system considered here, and it stresses the care that must be taken when using particle methods for kinetic equations.
\end{abstract}

\section{Introduction}

Often, the mean free path for a binary Coulomb interaction between two charged particles in the plasma that makes up the vast majority of the luminous universe is not small compared to the dynamical scales of the astrophysical system.
The infrequency of collisions in these astrophysical systems necessitates a kinetic description.
Kinetic equations are thus of critical importance to our understanding of astrophysical phenomena. 
However, the numerical integration of kinetic equations presents a significant challenge, because we require a representation of the solution in a six dimensional phase space, three position and three velocity variables, as well as in time.

It is unequivocally true that the most common numerical techniques for the solution of kinetic equations are Monte Carlo methods.
For the plasma kinetic equation, or Vlasov equation, this approach discretizes the particle distribution function as ``macroparticles,'' particles with some shape function. 
These macroparticles are then advanced along their characteristics and sampled to construct the required velocity moments to couple the plasma dynamics to the electromagnetic fields via Maxwell's equations \citep{Dawson:1962, Langdon:1970, Dawson:1983, birdsallbook, Lapenta:2012}.
This numerical method is traditionally called the particle-in-cell (PIC) method, since the results of the velocity moment computations are deposited onto the grid that discretizes Maxwell's equations.

The PIC method has been historically very fruitful and the core method of several production-level computational tools for simulating kinetic plasmas \citep[e.g.,][]{Fonseca:2008,Bowers:2009,Germaschewski:2016}.
However, because of its Monte Carlo nature, the PIC method introduces counting noise into the solution of the kinetic equation.
This numerical noise can manifest as a combination of numerical collisions and heating of the underlying distribution of particles and has been quantified in a number of studies \citep{Hockney:1968, Okuda:1970, Okuda:1972, Hockney:1971, Langdon:1979, Krommes:2007}.

While numerical heating can be quantified and controlled, the pollution of the solution with noise can have larger effects on the dynamics of the plasma.
For example, \citet{Camporeale:2016} have demonstrated that a large number of particles-per-cell is required to correctly identify wave-particle resonances and compare well with linear theory. 
More egregious examples include the counting noise induced transport found to be the source of disagreement between PIC models of turbulent transport in nuclear-fusion relevant simulations and corresponding continuum models of the same flavor of turbulence \citep{Nevins:2005}.
Here, a continuum model refers to a numerical model of a kinetic equation which directly discretizes the quantity of interest, the particle distribution function, on a phase space grid, directly solving a six-dimensional partial differential equation in time.

There are many means of reducing this noise.
The simplest strategy is to use more particles, but the counting noise decreases like $1/\sqrt{N_{ppc}}$, where $N_{ppc}$ is the number of particles per grid cell, and this slow convergence with increasing particle count can make the desired phase space accuracy prohibitively expensive computationally.
Other techniques explicitly modify the algorithm, such as the delta-f PIC method \citep{Parker:1993, Hu:1994, Denton:1995, Belova:1997, ChengJianhua:2013, Kunz:2014b}, very high order and more sophisticated particle shape functions, e.g., particle-in-wavelets \citep{vanyenNguyen:2010, vanyenNguyen:2011} and von Mises distributions based on Kernel Density Estimation theory \citep{Wu:2018}, and time-dependent deformable shape functions for the particles \citep{Coppa:1996}, the latter of which is an active area of research for the N-body Monte Carlo method applied to gravitational systems \citep{Abel:2012, Hahn:2015} and has recently been extended to PIC \citep{KatesHarbeck:2016}.

However, many of these modifications have their own deficiencies.
The delta-f PIC method can break down if the distribution function deviates significantly from its initial value, and the modifications to the particle shape introduce significant computational complexity to the algorithm.
This additional computational complexity makes application of these techniques to general kinetic systems more challenging, and preliminary work is focused on lower dimensional systems \citep{vanyenNguyen:2011} and post-processing analysis \citep{Totorica:2018}.
While the application of advanced particle shapes to even just the analysis of simulations pays dividends in reducing the noise in the solution \citep{Totorica:2018}, any issues due to noise that arise during the course of a simulation are not mitigated.

In this paper, we document an instance of disagreement in the underlying dynamics of competing plasma instabilities when studied with a PIC simulation and continuum kinetic simulation, and trace the origin of the disagreement to the counting noise introduced to the solution by the PIC algorithm.
We emphasize that the disagreement stemming from particle noise manifests not simply as numerical heating, but as a fundamental difference in the final state of the plasma’s nonlinear evolution.
The noise inherent to the PIC method leads to a sampling error in the computation of the current density, particularly for small numbers of particles-per-cell, and thus artificially introduces saturated small-scale magnetic field structure, while continuum simulations find that the magnetic field is strongly damped.
Particle noise is confirmed as the result of this disagreement with a combination of larger particle count simulations and post-process filtering of the saturated state.

We are inspired by the recently reported disagreement between the continuum kinetic simulations performed in \citet{Skoutnev:2019} and past PIC calculations in a similar parameter regime \citep{Kato:2008} of the competition of a class of Weibel-type instabilities driven by counter-streaming beams of plasmas \citep{Weibel:1959, Bornatici:1970, Davidson:1972}.
These Weibel-type instabilities are an interesting class of plasma instabilities, serving as a possible explanation for the observed magnetic fields in gamma ray bursts \citep{Medvedev:1999} and pulsar wind outflows \citep{Kazimura:1998}, and as a potential source of the seed magnetic field in a cosmological context \citep{Schlickeiser:2003, Lazar:2009}.
Importantly, while these Weibel-type instabilities robustly grow a magnetic field in the relativistic context, \citet{Skoutnev:2019} found that the non-relativistic limit of these instabilities was more complex, with a spectrum of unstable modes all having comparable growth rates vying for dominance.
While these Weibel-type instabilities are generally driven by counter-streaming beams of both protons and electrons, in this paper we will focus on the electron-driven variants of these instabilities.

In this regard, for the electron-driven, equal beam density and equal beam temperature, variants of these instabilities, two ratios primarily affect which mode grows the fastest: $u_d/c$, how non-relativistic the drift velocity of the beams is, and $v_{th_e}/u_d$, how much of the initial beam energy is internal energy versus kinetic energy.
Here, $u_d$ is the drift speed of each beam of electrons, $c$ is the speed of light, and $v_{th_e} = \sqrt{k_B T_e/m_e}$ is the electron thermal velocity.
In the relativistic case, $u_d \approx c$, the filamentation instability \citep{Fried:1959} is the fastest growing mode, and previous studies find robust magnetic field growth and saturation \citep{Fonseca:2003, Silva:2003, Nishikawa:2003, Nishikawa:2005, Kumar:2015, Takamoto:2018}.
Likewise, as the drift velocity becomes non-relativistic, so long as the beams are ``hot,'' i.e., $v_{th_e} \approx u_d$, the disruption of the fast-growing two-stream instability leads to a secondary Weibel instability.
The disruption of the saturated two-stream modes by the more slowly growing filamentation instability leads to energy conversion dominantly in only one velocity dimension, since the two-stream instability is one-dimensional, and this temperature anisotropy provides another source of free energy for the Weibel instability and thus a means of supporting a saturated magnetic field \citep{Schlickeiser:2003, Kato:2008}. 
The saturated magnetic field from the disruption of two-stream modes leads to the same levels of magnetization as previous studies of the Weibel instability and filamentation instability in isolation, in either one dimension or two dimensions \citep{Morse:1971, Califano:1997, Califano:1998, Cagas:2017}.

These results stand in stark contrast to the findings of \citet{Skoutnev:2019} as the ratio of $v_{th_e}/u_d$ is decreased further and the inter-penetrating beams are made ``colder.''
As the temperature of the beams is reduced, the growth rates of a spectrum of oblique modes increase, modes which arise due to perturbations between parallel (two-stream) and perpendicular (filamentation) to the drift direction \citep{Bret:2009}.
These hybrid two-stream-filamentation modes begin to have comparable growth rates to the two-stream instability---see Figure 1 in \citet{Skoutnev:2019}.
These fast growing oblique modes rapidly deplete the free energy in the inter-penetrating flows, eliminating the channel for magnetic field growth via a combination of the more slowly growing filamentation instability and secular Weibel instability from the residual temperature anisotropy of saturated two-stream modes studied in, e.g., \citet{Kato:2008}.

With the free energy depleted, the oblique modes are then able to collisionlessly damp on the electrons.
This collisionless damping converts the electromagnetic energy from the saturated oblique modes to electron thermal energy, leaving little residual magnetic energy in the system, along with a highly structured distribution function in phase space due to the mixing of nonlinearly saturated oblique modes.
This additional phase space structure serves as an added marker for the collapse of the magnetic field via damping of the oblique modes.

The collapse of the magnetic field in the non-relativistic, cold limit in the continuum kinetic simulations presented in \citet{Skoutnev:2019} had not been previously reported, and in fact disagreed with the PIC results of \citet{Kato:2008} in the ``cold'' parameter regime, $u_d/c = 0.1, v_{th_e}/u_d = 0.1$.
\citet{Skoutnev:2019} identified a number of potential explanations for this disagreement, such as the shock geometry---\citet{Kato:2008} performed Weibel-mediated collisionless shock simulations while \citet{Skoutnev:2019} considered only perturbations to an initially unstable system and not a driven system such as a collisionless shock.
In addition, \citet{Skoutnev:2019} focused solely on the electron dynamics with the protons forming a neutralizing background while \citet{Kato:2008} included the dynamics of the protons in their collisionless shock simulations.
We consider the effect of the difference in numerical method in this paper and perform identical simulations to \citet{Skoutnev:2019} with a PIC method, focusing on the electron-only variants of these Weibel-type instabilities with an initial-value problem. 

\section{Methods and Results}

We concern ourselves with the evolution of the Vlasov equation of a species $s$,
\begin{align}
    \frac{\partial f_s}{\partial t} + \mathbf{v} \cdot \nabla f_s + \frac{q_s}{m_s}\left( \mathbf{E} + \mathbf{v} \times \mathbf{B} \right) \cdot \nabla_{\mathbf{v}} f_s = 0,
\end{align}
coupled to Maxwell's equations via self-consistent currents. 
The Vlasov--Maxwell system of equations is numerically integrated with the PIC code \pic~\citep{Zeiler:2002} and the \gke~simulation framework, which contains a continuum Vlasov--Maxwell solver \citep{Juno:2018}.
Note that the \gke~simulation framework also contains a Fokker--Planck collision operator, and can thus numerically integrate the Vlasov--Maxwell--Fokker--Planck system of equations \citep{Hakim:2019}.

We initialize an electron-proton plasma in two spatial dimensions and two velocity dimensions (2X2V), with two, uniform, equal density, counter-streaming, Maxwellian beams of electrons,
\begin{align}
f_{0,e}(v_x,v_y)=\frac{n_0}{2\pi v_{th_e}^2}e^{-\frac{v_x^2}{2v_{th_e}^2}}\left[e^{-\frac{(v_y-u_d)^2}{2v_{th_e}^2}}+e^{-\frac{(v_y+u_d)^2}{2v_{th_e}^2}}\right].
\end{align}
The protons form a stationary, neutralizing background. 
Here, $n_0 = 1$ is a density normalization.
To excite the zoo of instabilities, two-stream, filamentation \citep{Fried:1959}, and electromagnetic oblique \citep{Bret:2009}, a collection of electric and magnetic fluctuations are initialized,
\begin{align}
    B_z(t=0) = \sum_{n_x, n_y = 0}^{16, 16} \tilde{B} \sin\left( \frac{2 \pi n_x x}{L_x} + \frac{2 \pi n_y y}{L_y} + \tilde{\phi}\right ),
\end{align}
with equivalent perturbations in $E_x$ and $E_y$. 
Here, $\tilde{B}$ and $\tilde{\phi}$ are random amplitudes and phases, with the random amplitudes chosen such that all three fields have equal initial average energy densities, $\langle \epsilon_0 E_x^2/2 \rangle = \langle \epsilon_0 E_y^2/2 \rangle = \langle B_z^2/2\mu_0 \rangle \approx 10^{-7} E_K$, where $E_K$ is the total initial energy of the two counter-streaming beams of electrons.

Further details of the simulations are as follows.
For all simulations, the box size is chosen based on the $u_d/c = 0.1, v_{th_e}/u_d = 0.1$ simulation in \citet{Skoutnev:2019}, $L_x = 2.7 d_e, L_y = 3.1 d_e$, where $d_e = c/\omega_{pe}$ is the electron inertial length and $\omega_{pe} = \sqrt{e^2 n_e/\epsilon_0 m_e}$ is the electron plasma frequency.
This box size is chosen to fit the faster growing filamentation mode, $k_{max}^{FI} = 2 \pi /L_x$, and an integer number of two-stream modes $k_{max}^{TS} = 2 \pi n/L_y$, while keeping the box size roughly square $L_x \approx L_y$.
The \gke~continuum Vlasov--Maxwell simulation is run with $N_x = N_y = 48$ grid points in configuration space, $v_{max} = [-3 u_d, 3 u_d]$ velocity space extents, $N_v = 64^2$ in the two velocity dimensions, and piecewise quadratic Serendipity elements \citep{Arnold:2011}---see \citet{Juno:2018} for details on the discontinuous Galerkin discretization of the Vlasov--Maxwell system of equations.
The \pic~simulations are run with configuration space resolution $\Delta x = \Delta y = 0.014 d_e$, and the number of particles per cell is varied between $N_{ppc} = [12, 120, 1200, 12000]$, with both linear and quadratic particle shapes.
In addition, the \pic~simulations employ Marder's correction \citep{Marder:1987} via a multigrid method to enforce charge conservation.
Periodic boundary conditions are employed in configuration space, and for only the continuum \gke~simulation, zero-flux boundary conditions are used in velocity space, along with a small amount of collisions, $\nu_{ee} = 4 \times 10^{-5} \omega_{pe}$, for velocity space regularization---see \citet{Hakim:2019} for details of the collision operator implementation in \gke.

\begin{figure}
    \centering
    \includegraphics[width=0.49\textwidth]{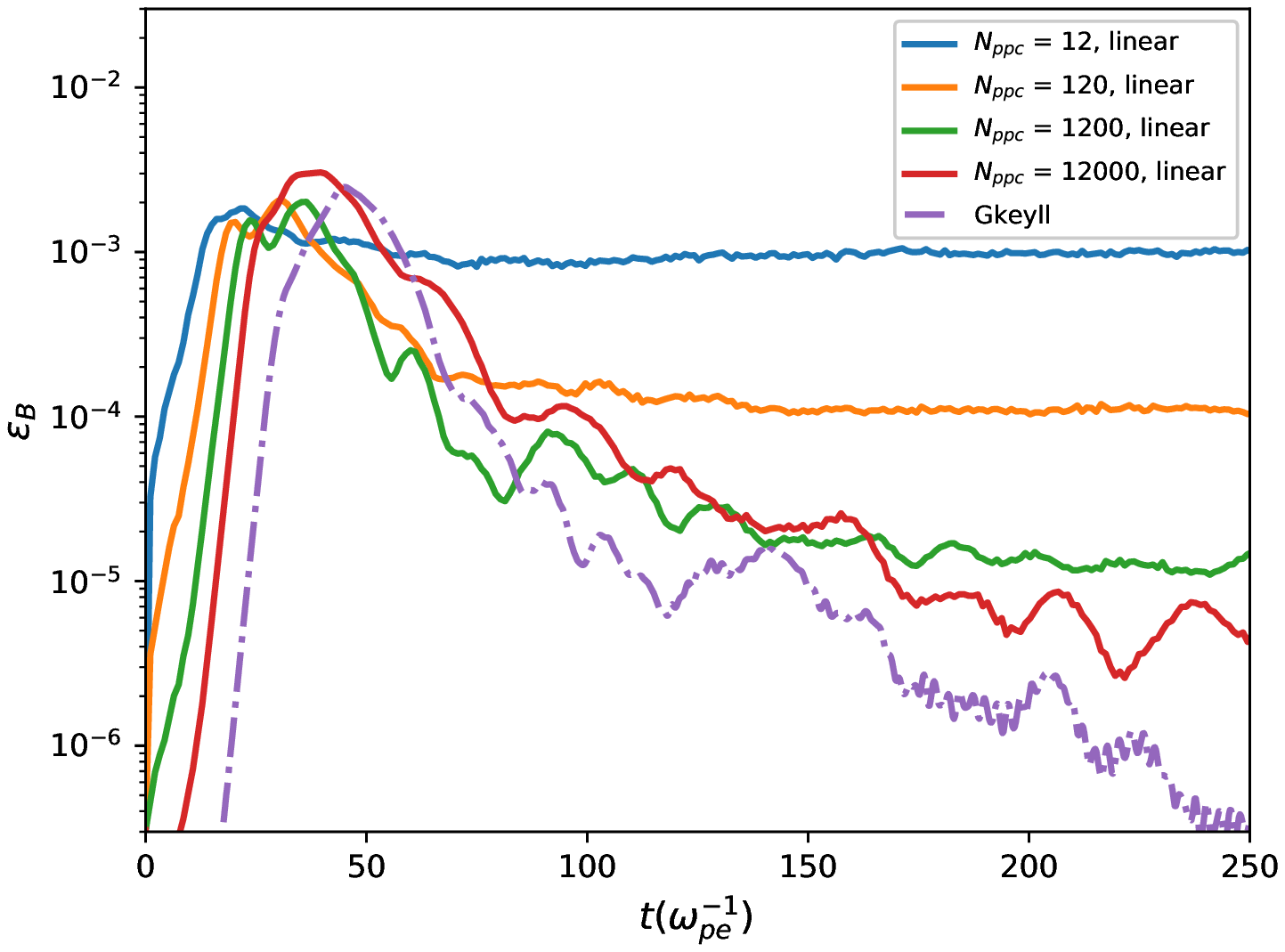}
    \includegraphics[width=0.49\textwidth]{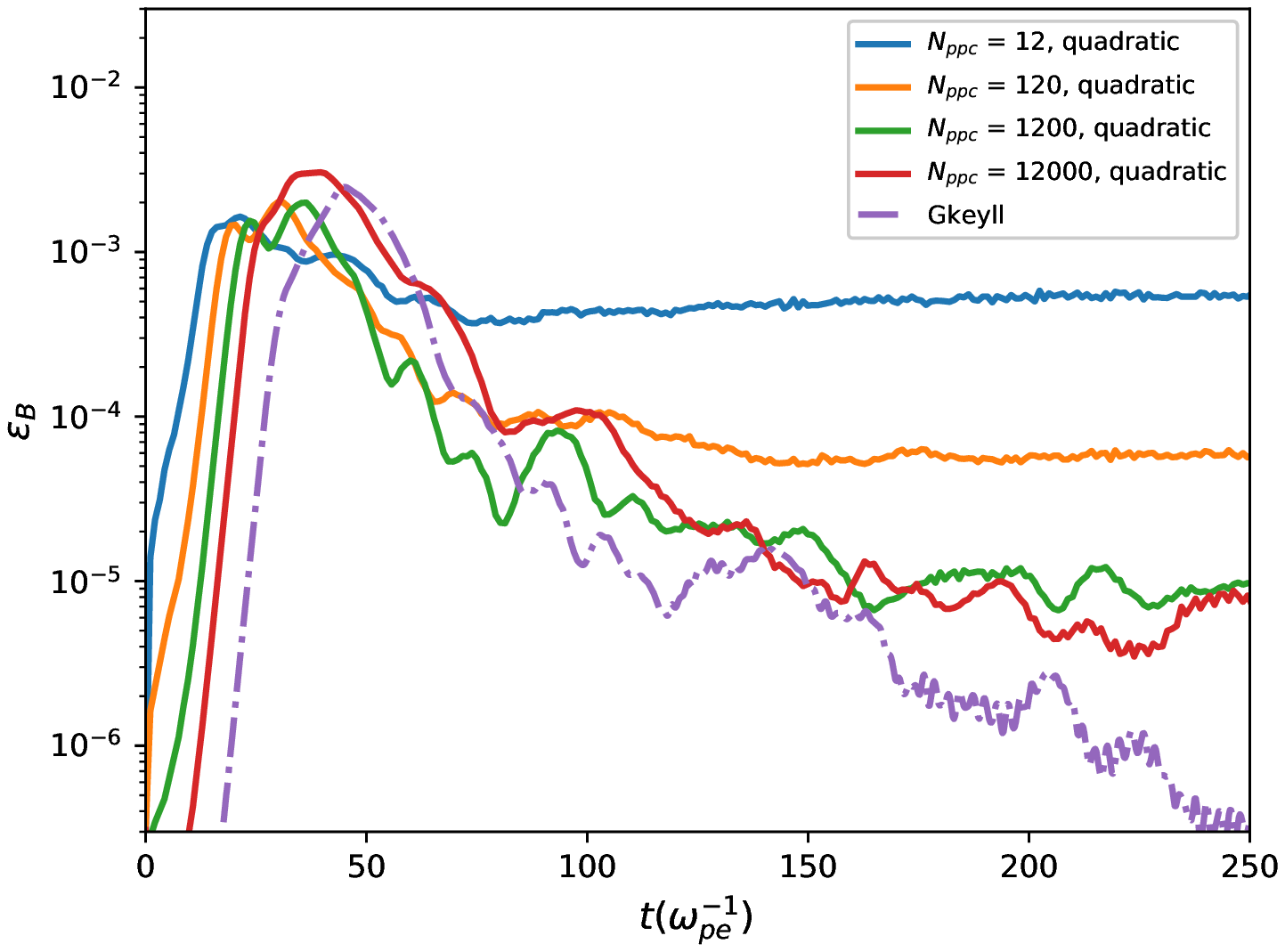}
    \caption{Evolution of the box-integrated magnetic field energy, $\epsilon_B = 1/(2 \mu_0) \int B_z^2$, normalized to the initial electron energy for a sequence of \pic~calculations varying particle number and particle shape and compared to a fiducial \gke~calculation. While there is initial growth of the magnetic field due to the electromagnetic component of the oblique modes, the oblique modes quickly damp on the electrons, leading to the decay of the magnetic field energy in the continuum \gke~simulation and high particle count \pic~simulations. With the free energy of the counter-streaming electron beams depleted by the fast-growing oblique modes, the continuum and more highly resolved PIC calculations cannot support any saturated magnetic field structure. However, with reduced particle count, the magnetic field energy saturates to a fixed value. Modest increases in the number of particles per cell and higher order particle shapes reduce the magnitude of the saturated magnetic field, but only with more substantial increases in particle count do we observe similar late time behavior between the \pic~PIC and \gke~continuum simulations.}
    \label{fig:eb-pic-gke}
\end{figure}

The evolution of the box-integrated magnetic energy from this suite of simulations is shown in Figure \ref{fig:eb-pic-gke}.
We can clearly identify the collapse of the magnetic field observed in \citet{Skoutnev:2019} in both the continuum \gke~simulations and the high particle count \pic~simulations.
However, as the number of particles-per-cell is decreased, the magnetic field attains a particle-per-cell dependent saturated state.
The saturation level has some sensitivity to the particle shape, with the quadratic spline particles saturating at a lower level than the linear spline particles.
But even the modestly high particle counts saturate at a still higher amplitude than what we expect from the continuum calculation and more resolved PIC calculations.

\begin{figure}
    \centering
    \includegraphics[width=\textwidth]{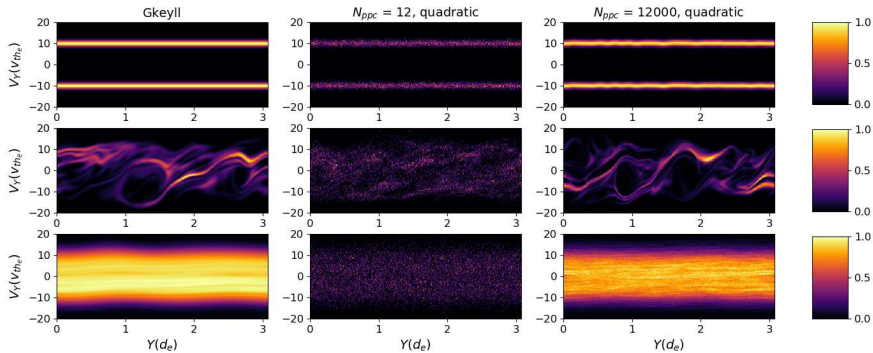}
    \caption{Evolution of the electron distribution function from \gke~(left column) and $N_{ppc}=12$ and $N_{ppc} = 12000$, quadratic spline, \pic~(middle and right column) simulations in $y-v_y$. The rows show the evolution of the distribution function in time, $t = 0 \omega_{pe}^{-1}$ (top row), the peak of the magnetic field energy, $t = 45 \omega_{pe}^{-1}$ for the \gke~simulation and $t = 35 \omega_{pe}^{-1}$ for the \pic~simulations (middle row), and the end of the simulation $t = 250 \omega_{pe}^{-1}$ (bottom row). These distribution function cuts are generated by integrating over a narrow slice of the interior of $L_x$ and all of $v_x$. Specifically, we integrate the \gke~simulation over the middle two grid cells, $\Delta x = 0.1 d_e$, and sample particles from the corresponding extent in the \pic~simulations. In addition, to generate the velocity space representation in $v_y$ of the \pic~simulations, the particles are binned into 101 equally space bins from $-20 v_{th_e}$ to $20 v_{th_e}$. All three simulations begin with good phase space resolution, as the initial particle distribution function can be sampled precisely even with only $N_{ppc} = 12$. However, we can see that the evolution of phase space is much less well-resolved with very few particles-per-cell, and because the nonlinear dynamics of the saturated instabilities lead to a phase space filling electron distribution function, the effective phase space resolution of the $N_{ppc} = 12$ calculation has decreased significantly by the end of the simulation.}
    \label{fig:yvycomp}
\end{figure}

\begin{figure}
    \centering
    \includegraphics[width=\textwidth]{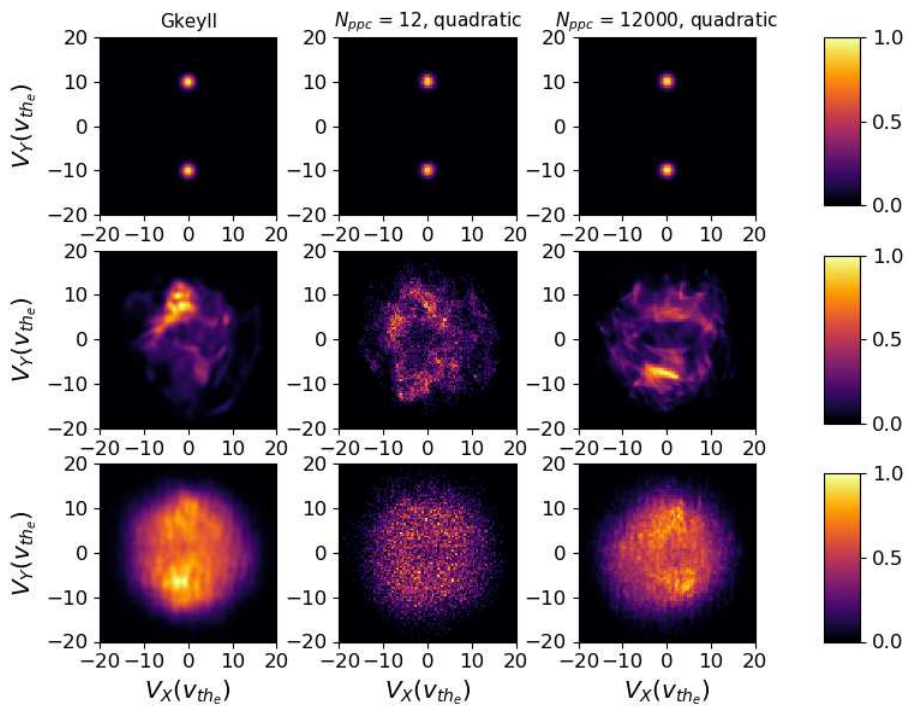}
    \caption{The same evolution of the electron distribution function shown in Figure \ref{fig:yvycomp} from \gke~(left column) and $N_{ppc}=12$ and $N_{ppc} = 12000$, quadratic spline, \pic~(middle and right column) simulations, but now in $v_x-v_y$. The rows show the evolution of the distribution function in time, $t = 0 \omega_{pe}^{-1}$ (top row), the peak of the magnetic field energy, $t = 45 \omega_{pe}^{-1}$ for the \gke~simulation and $t = 35 \omega_{pe}^{-1}$ for the \pic~simulations (middle row), and the end of the simulation $t = 250 \omega_{pe}^{-1}$ (bottom row). These distribution function cuts are generated by integrating over a narrow slice of the interior of $L_x$ and all of $y$. As in Figure \ref{fig:yvycomp}, we integrate the \gke~simulation over the middle two grid cells, $\Delta x = 0.1 d_e$, and sample particles from the corresponding extent in the \pic~simulations. In addition, to generate the velocity space representation in $v_x$ and $v_y$ of the \pic~simulations, the particles are binned into 101 equally space bins from $-20 v_{th_e}$ to $20 v_{th_e}$ for both velocity dimensions. Again, the phase space resolution at the beginning of the simulation is acceptable, even with only $N_{ppc} = 12$, because we can sample particles efficiently for the cold distribution. However, because the unstable oblique modes efficiently convert this free energy into electromagnetic energy and then back into electron thermal energy, the final electron distribution fills a much larger volume of phase space and the effective phase space resolution has plummeted by the end of the simulation.}
    \label{fig:vxvycomp}
\end{figure}

To understand this pseudo-saturation of the magnetic field, we show the phase space structure from the \gke~simulation and a subset of the \pic~simulations in Figures \ref{fig:yvycomp} and \ref{fig:vxvycomp}.
We plot the results from the \gke~simulation (left column), along with the $N_{ppc} = 12$ and $N_{ppc} = 12000$, quadratic spline particle shape, \pic~simulations (middle and right columns) in $y-v_y$ (Figure \ref{fig:yvycomp}) and $v_x-v_y$ (Figure \ref{fig:vxvycomp}).
The three rows denote the beginning of the simulation, $t = 0 \omega_{pe}^{-1}$ (top row), peak of the magnetic field energy, $t = 45 \omega_{pe}^{-1}$ for the \gke~simulation and $t = 35 \omega_{pe}^{-1}$ for the \pic~simulations (middle row), and end of the simulation, $t = 250 \omega_{pe}^{-1}$ (bottom row).
These distribution function cuts are generated from the full 2X2V phase space by integrating over the interior $0.1 d_e$ in the $x$ dimension, i.e., the middle two grid cells in $x$ in the \gke~simulation, and the equivalent $x$ extents in the \pic~simulations, along with an integration in the entirety of $v_x$ for the $y-v_y$ cut in Figure \ref{fig:yvycomp}, and the entirety of $y$ for the $v_x-v_y$ cut in Figure \ref{fig:vxvycomp}.
We note that to generate the velocity space representation in the \pic~simulations, the particles are binned into 101 equally spaced bins from $-20 v_{th_e}$ to $20 v_{th_e}$ in $v_y$ for Figure \ref{fig:yvycomp} and both $v_x$ and $v_y$ for Figure \ref{fig:vxvycomp}.

We observe that, regardless of the ``coarseness'' of our particle resolution, the simulations begin with good phase space resolution.
Although the beams are cold and the drift speed is large, $u_d \gg v_{th_e}$, even with $N_{ppc} = 12$, we can concentrate phase space resolution in the initial distribution function for the \pic~calculations.
However, as the instabilities evolve and lead to the electrons filling phase space, the low particle count simulation in Figures \ref{fig:yvycomp} and \ref{fig:vxvycomp} becomes much less well resolved, with an inability to distinguish the features observed at saturation in the \gke~and $N_{ppc} = 12000$ simulations, and the steady-state isotropic distribution becoming poorly sampled.

To understand the effect this decreased effective phase space resolution can have on the dynamics of the plasma in the $N_{ppc} = 12$ \pic~simulation, consider Amp\`ere's Law in steady state,
\begin{align}
    \nabla_{\mathbf{x}} \times \mathbf{B} = \mu_0 \mathbf{J} = \mu_0 \sum_s q_s \int \mathbf{v} f_s \thinspace d\mathbf{v},
\end{align}
where we have assumed $\partial \mathbf{E}/\partial t \rightarrow 0$ and substituted the charge-weighted sum over each species' first velocity moment for the current density.
While this computation of the current is straightforward for the grid-based continuum method in \gke, for the \pic~PIC calculations this velocity moment is given by the sum over each individual macroparticle's velocity
\begin{align}
    \sum_s q_s \int \mathbf{v} f_s \thinspace d\mathbf{v} = \sum_s q^{PIC}_s \sum_{j=1}^N \mathbf{V}_{s,j} S (\mathbf{x} - \mathbf{X}_{s,j}),
\end{align}
where $\mathbf{X}_{s,j}$ and $\mathbf{V}_{s,j}$ are the $j$ macroparticle positions and velocities for species $s$, $q^{PIC}_s$ is the charge of the macroparticle of species $s$, and $S(\mathbf{x})$ is the shape function in configuration space.

Because the current density is computed from discrete macroparticles, if any fluctuations in the current density are uncorrelated on some spatial scale, we expect these fluctuations can give rise to a magnetic field.
This behavior would be akin to the ``quasi-thermal,'' electrostatic noise which arises from uncorrelated density fluctuations in a plasma, both in real plasma systems \citep{MeyerVernet:1986} and in PIC calculations \citep{Langdon:1979}.
We expect, in analogy with the Debye length for electrostatic shielding and density fluctuations, the electron skin depth, $d_e$, is the scale below which current density fluctuations will not be effectively shielded by the magnetic field.
Since these fluctuations are uncorrelated, their ensemble average, i.e., their average over many realizations of the plasma system, will be zero, but their root-mean-square (RMS) average might not be zero.
In other words, the current density computed from the collection has zero ensemble average,
\begin{align}
    \left \langle \mathbf{J}^{PIC}  \right \rangle = \sum_s q^{PIC}_s \left \langle \sum_j \mathbf{V}_{s,j} S (\mathbf{x} - \mathbf{X}_{s,j}) \right \rangle = 0,
\end{align}
where $\langle \cdot \rangle$ denotes the ensemble average, but the squared RMS,
\begin{align}
    \left |\mathbf{J}_{RMS}^{PIC} \right |^2 =  \sum_s \frac{1}{N^{PIC}}\left (q^{PIC}_s \right)^2 \sum_j \sum_k \mathbf{V}_{s,j} S (\mathbf{x} - \mathbf{X}_{s,j}) \mathbf{V}_{s,k} S (\mathbf{x} - \mathbf{X}_{s,k}),
\end{align}
where $N^{PIC}$ is the number of PIC particles being summed over, may have non-zero ensemble average,
\begin{align}
    \left \langle \left | \mathbf{J}_{RMS}^{PIC} \right |^2 \right \rangle =  \sum_s \frac{1}{N^{PIC}} \left ( q^{PIC}_s \right)^2 \left \langle \sum_j \sum_k \mathbf{V}_{s,j} S (\mathbf{x} - \mathbf{X}_{s,j}) \mathbf{V}_{s,k} S (\mathbf{x} - \mathbf{X}_{s,k}) \right \rangle \neq 0.
\end{align}

If we assume the current is carried solely by the electrons and consider a volume of $\sim d_e^2$ in two dimensions, we can make a back-of-the-envelope calculation for the size of the RMS current density,
\begin{align}
    \left \langle \left | \mathbf{J}_{RMS}^{PIC} \right |^2 \right \rangle & \sim \frac{ \left ( e^{PIC} \right )^2}{ \left ( d_e^{PIC} \right )^4} \frac{1}{N^{PIC}} \sum_j \left \langle |\mathbf{V}_j|^2 \right \rangle, \notag \\
    & \sim \frac{ \left ( e^{PIC} \right )^2}{ \left ( d_e^{PIC} \right )^4} N^{PIC} \left (v^{PIC}_{th_e} \right )^2, \label{eq:currentrms}
\end{align}
where we have added the superscript $PIC$ to all quantities, charge, $e^{PIC}$, electron inertial length, $d_e^{PIC}$, and electron thermal velocity $v_{th_e}^{PIC}$ to emphasize these are the macroparticle quantities.

In general, because a macroparticle represents many particles in the real plasma system, all intrinsic properties, e.g., the mass and charge, must be scaled by the appropriate macroparticle factor when comparing a quantity computed with the PIC method to the actual physical quantity.
Comparing the RMS current density from the PIC method to the physical RMS current density from a fiducial plasma, we have
\begin{align}
    \frac{\left \langle \left | \mathbf{J}_{RMS}^{PIC}  \right |^2 \right \rangle}{\left \langle \left | \mathbf{J}_{RMS} \right |^2 \right \rangle} = \frac{\frac{\left ( e^{PIC} \right )^2}{\left ( d_e^{PIC} \right )^4} N^{PIC} \left (v^{PIC}_{th_e} \right )^2}{\frac{e^2}{d_e^4} N v_{th_e}^2},
\end{align}
where $N$ is the number of particles in the real plasma system being modeled.
Since every charge is a macroparticle charge, the difference between the macroparticle charge and the elementary charge is
\begin{align}
    \frac{e^{PIC}}{e} = \frac{F^{macro} e}{e} = \frac{N}{N_{ppc} N_{cells}},
\end{align}
where $F^{macro}$ is the scaling factor for the number of particles a macroparticle represents and $N_{cells}$ is the number of grid cells we have summed over to construct our $d_e^2$ volume.
In other words, when $N_{ppc} N_{cells} < N$, the charge in the PIC method must be scaled by the appropriate factor since the macroparticle is representing some (potentially large) number of particles.

We can proceed in a similar fashion for the other terms.
We note that the electron plasma frequency, $\omega_{pe} = \sqrt{e^2 n/\epsilon_0 m_e}$, is macroparticle independent because the macroparticle factors cancel in the charge, density, and mass.
Therefore, the electron inertial length, $d_e = c/\omega_{pe}$, is also macroparticle independent.
To determine the factor for the electron thermal velocity, we note that substitution of one factor of $d_e^2$ reveals that
\begin{align}
    \frac{\left (v_{th_e}^{PIC} \right )^2}{d_e^2} = \frac{\left (v_{th_e}^{PIC} \right )^2}{c^2} \omega_{pe}^2,
\end{align}
and since the ratio of the electron thermal velocity to the speed of light is independent of macroparticle size, we have no additional factor from the electron thermal velocity.
We are thus left with
\begin{align}
    \frac{\left \langle \left |\mathbf{J}_{RMS}^{PIC} \right |^2 \right \rangle}{\left \langle \left | \mathbf{J}_{RMS} \right |^2 \right \rangle} = \frac{N}{N_{ppc} N_{cells}},
\end{align}
after substitution of $N^{PIC} = N_{ppc} N_{cells}$.
We can then massage Eqn.~(\ref{eq:currentrms}) to a more general form with the appropriate scaling factor,
\begin{align}
    \left \langle \left |\mathbf{J}_{RMS}^{PIC} \right |^2 \right \rangle \sim \left (\frac{N}{N_{cells} N_{ppc}} \right ) \frac{e^2}{d_e^2} n v^2_{th_e}, \label{eq:picrmscurrent}
\end{align}
where $n$ is the number density of the plasma.
In the limit that every macroparticle represents a single charged particle in the real plasma system, Eqn.~(\ref{eq:picrmscurrent}) would reduce to the estimate of the continuum RMS current density spectrum from uncorrelated current density fluctuations. 

Substitution of the estimate in Eqn.~(\ref{eq:picrmscurrent}) into Amp\`ere's law gives us
\begin{align}
    \frac{1}{2\mu_0} \left \langle \left |\mathbf{B}_{RMS}^{PIC} \right |^2 \right \rangle & \sim \frac{d^2_e \mu_0}{2} \left \langle \left | \mathbf{J}_{RMS}^{PIC} \right |^2 \right \rangle, \notag \\
    & \sim \left (\frac{N}{N_{cells} N_{ppc}} \right ) \frac{d^2_e \mu_0}{2} \frac{e^2}{d_e^2} n v^2_{th_e},
\end{align}
which, upon rearrangement of the expression using the fact that $c^2 = 1/\epsilon_0 \mu_0$ and our definitions of the electron thermal velocity, $v_{th_e}$, the electron plasma frequency, $\omega_{pe}$, and electron skin depth, $d_e$, we obtain
\begin{align}
    \frac{1}{2\mu_0} \left \langle \left | \mathbf{B}_{RMS}^{PIC} \right |^2 \right \rangle & \sim \frac{1}{2} \frac{N}{N_{cells} N_{ppc}} \frac{T}{d_e^2} \notag \\
    & \sim \frac{1}{2} \frac{n T}{N_{cells} N_{ppc}}. \label{eq:bPICSpectrum}
\end{align}
This estimate is similar to \citet{Tajima:1992} for the continuum noise spectrum of current fluctuations, but with the additional $N_{cells} N_{ppc}$ term for how the fluctuation spectrum scales with the number of particles-per-cell.
However, this calculation is fundamentally different from the \citet{Tajima:1992} computation of the continuum noise spectra because we expect the fluctuation spectrum to decrease with increasing $N_{ppc}$, as we are estimating an anomalous source of fluctuations.
The PIC method is still a numerical discretization of the Vlasov equation, which in the absence of true discrete particle effects such as collisions should not have a fluctuation spectrum due to uncorrelated current density fluctuations.

We note that the magnitude of the magnetic field fluctuations in Eqn.~(\ref{eq:bPICSpectrum}) inversely depends on the number of particles-per-cell, as we expect from the scaling of the saturated magnetic field amplitude in Figure~\ref{fig:eb-pic-gke}, where we observe the saturation level decreasing by roughly an order of magnitude when we increase the number of particles by a factor of ten. 
Importantly, the magnitude of the magnetic field fluctuations also depends on the temperature, thus explaining the observed saturated magnetic field state in the smaller $N_{ppc}$ \pic~simulations.
Despite a reasonable sampling accuracy at the beginning of the simulation for the initially cold beams, the large electron temperature increase from the oblique mode dynamics leads to a larger noise-floor of magnetic field fluctuations.
In other words, the number of particles per cell, $N_{ppc}$, is fixed, so increases in the temperature of the distribution being sampled inevitably increases the fluctuation amplitude arising from noise.
By the end of the simulation, we are much more poorly sampling the saturated electron distribution function because of the distribution's increased temperature, and we thus obtain a saturated magnetic field from the resulting RMS fluctuations in the current density.
The final, much hotter, electron distribution would require higher phase space resolution, i.e., a larger number of particles-per-cell, to effectively represent and reduce the RMS current density fluctuations which arise from the noise inherent to the PIC method.

\begin{figure}
    \centering
    \includegraphics[width=0.49\textwidth]{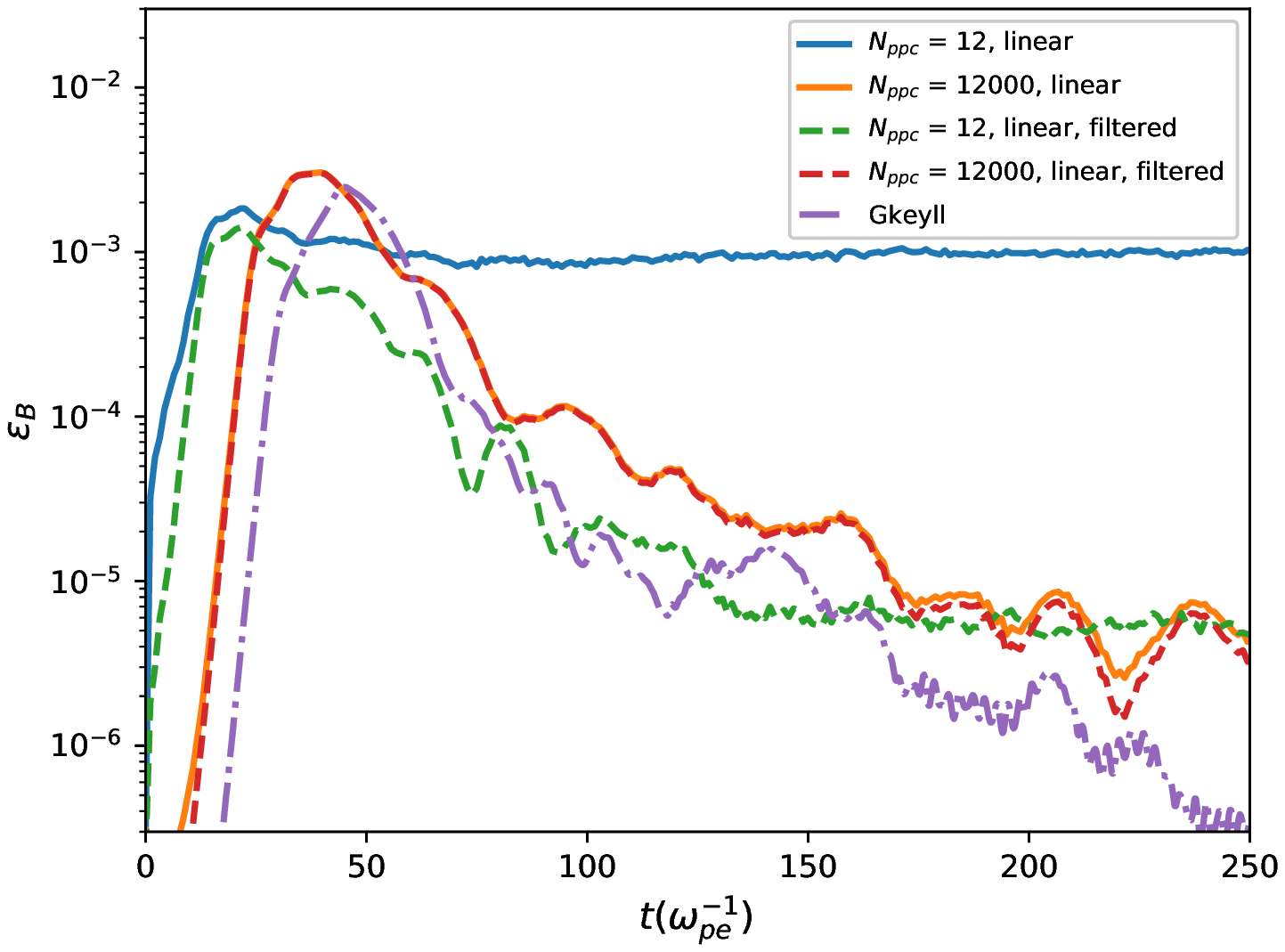}
    \includegraphics[width=0.49\textwidth]{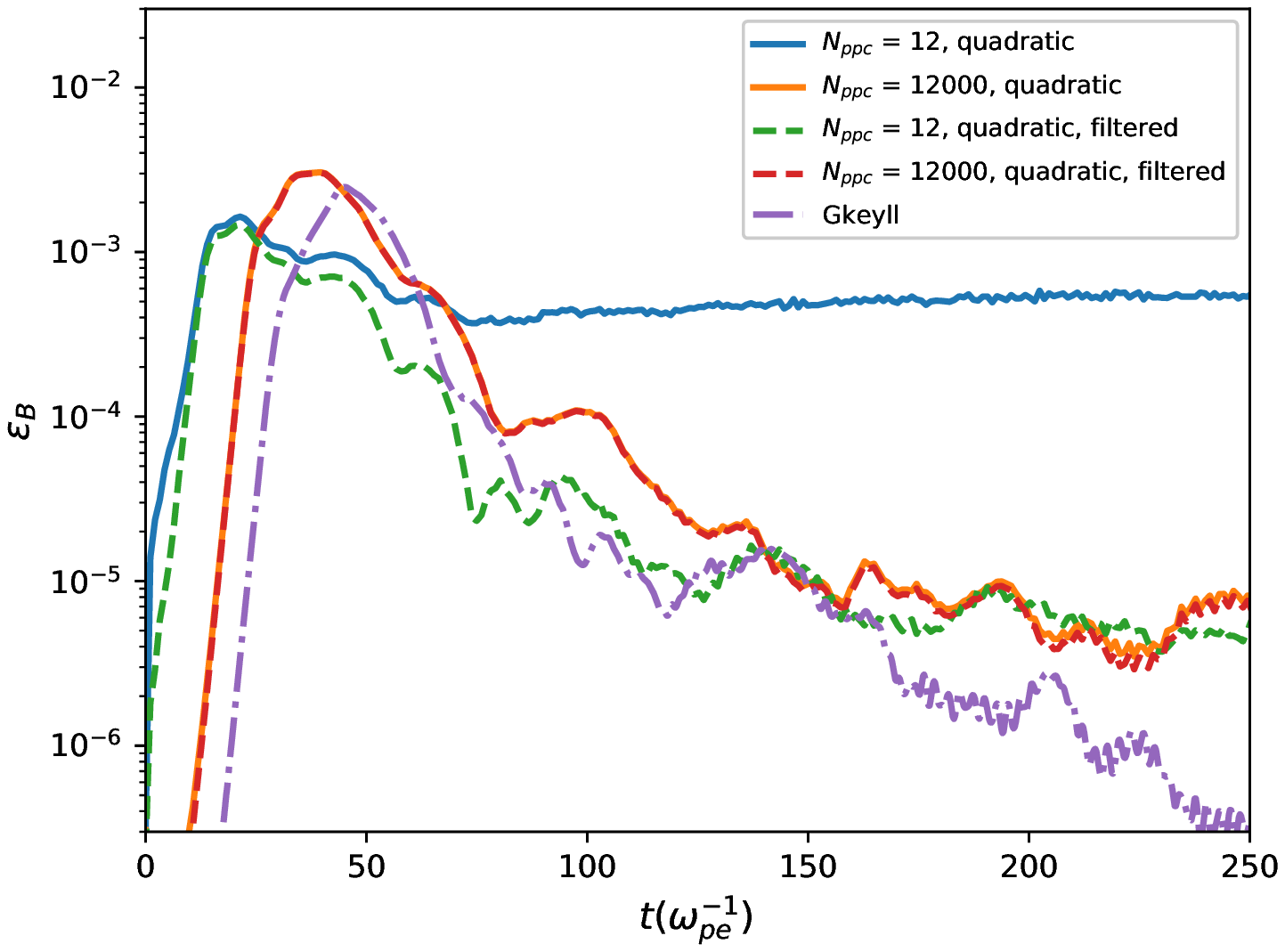}
    \caption{Evolution of the box-integrated magnetic field energy, $\epsilon_B = 1/(2 \mu_0) \int B_z^2$, normalized to the initial electron energy, with a $3 \times 3$ spatial boxcar filter applied in post-processing to the $N_{ppc} = 12$ and $N_{ppc} = 12000$ \pic~simulations, with the corresponding unfiltered \pic~data and the \gke~simulation for reference. The use of a filter on small spatial scales assists in the smoothing of the noise-generated magnetic field for this initial value problem, though the magnitude of the magnetic field collapse still does not agree with the fiducial \gke~simulation.}
    \label{fig:eb-filtered}
\end{figure}

It is natural to ask if this noise-generated magnetic field can be filtered in some fashion to restore the solution to the results found in the continuum \gke~simulation and converged \pic~simulations.
We plot in Figure \ref{fig:eb-filtered} the result of a boxcar smoothing on a $3 \times 3$ stencil in configuration space, i.e., a spatial average over the smallest scales in the \pic~simulation.
Clearly, the averaging over the small spatial scales assists in the restoration of the collapse of the magnetic field, even if the \pic~calculations still do not reproduce the exact quantitative level of collapse as the \gke~simulations.
We emphasize that the filtering procedure performed here is done in post-processing, and while this post-processing procedure works here for this initial value problem, we caution that for a driven system such as a Weibel-mediated collisionless shock, \emph{in situ} filtering may be required to restore the results of instability collapse, lest a pseudo-saturated magnetic field pollute the dynamics of the driven system.
But, aggressive \emph{in situ} filtering should also be employed with caution, as it is the saturated oblique modes that lead to the magnetic field collapse, and filtering \emph{in situ} may average over the oblique mode dynamics and prevent the collisionless damping of the electromagnetic fluctuations.

\section{Discussion and Summary}

We conclude having demonstrated the role particle noise can play in the dynamics of the electron-only variants of Weibel-type instabilities. 
In the non-relativistic, cold limit, we should find no net magnetization from the free energy of the initial counter-streaming electron beams due to the dynamically important electromagnetic oblique modes depleting this free energy and then damping the magnetic field, but a particle-per-cell dependent, saturated magnetic field can arise in underresolved PIC simulations due to noise-generated fluctuations in the current density.
This result on its own suggests a re-examination of some Weibel-relevant calculations may be necessary to accurately assess the role of these electromagnetic oblique modes.
The noise-driven saturated magnetic field in the low $N_{ppc}$ simulations is motivating for collisionless shock simulations with the PIC method, which often employ $N_{ppc} = 10-100$ \citep[see, e.g.,][]{Spitkovsky:2005, Kato:2008, Fiuza:2012, Huntington:2015}.
Beyond a re-exploration of Weibel-mediated collisionless shocks in the parameter regime considered here, studies of the Biermann battery \citep{Biermann:1950}, another candidate for the development of a magnetic field in the very early universe, have found the presence of the electron Weibel instability as one pushes to larger system size \citep{Schoeffler:2014, Schoeffler:2016, Schoeffler:2018}.
Thus, a study of the Biermann battery in this ``cold'' parameter regime is both timely due to the discovery of the importance of the oblique mode dynamics by \citet{Skoutnev:2019} and likely requires care given the results presented here on the sensitivity of the oblique mode dynamics to phase space resolution.

We do not wish to appear antagonistic towards the PIC method.
There are applications which historically have used fewer particles-per-cell, and our study suggests the physics of oblique modes requires higher effective phase space resolution.
In general, the phase space resolution required for a kinetic simulation to adequately represent the physics of interest is difficult to know \emph{a priori} \citep[see, e.g.,][]{Barnes:2010}.
It is not necessarily surprising that certain phenomena may require higher phase space resolution.
In this regard, the continuum method employed here should be viewed as a complementary approach to the general task of modeling kinetic plasmas because of the high resolution the method provides in phase space at an acceptable computational cost---the $N_{ppc} = 1200$ linear spline \pic~simulation has roughly the same cost as the $48^2 \times 64^2$, piecewise quadratic Serendipity element \gke~simulation.
We believe the power of this complementary continuum kinetic approach is made manifest by the results of this study, in addition to previous studies of similar phenomena in the unmagnetized regime, such as collisionless shocks \citep{Pusztai:2018, Sundstrom:2019}, the one-dimensional Weibel instability \citep{Cagas:2017}, and the plasma dynamo \citep{Pusztai:2020}. 

We wish to emphasize that the results of this study alone do not yet answer the question of the source of disagreement between \citet{Skoutnev:2019} and \cite{Kato:2008} in the non-relativistic, cold parameter regime.
On one hand, particle noise appears to be a possible component of the disagreement, with the potential for a noise-generated magnetic field to modify the dynamics in a driven system such as a Weibel-mediated collisionless shock.
On the other hand, this comparison is incomplete until the effects of the protons and shock geometry are also considered.
A recent study \citep{Matteucci:2019}, found that the growth of the magnetic field in laser ablation simulations matched solely the ion Weibel theory, as opposed to the combined ion-electron Weibel theory.
In this non-relativistic parameter regime, it is possible that the protons, or more generally any ion species, are solely responsible for magnetic field growth, and the protons receive no assistance from the electron Weibel instability due to the oblique mode dynamics of the electrons.
A further study of the phase space dynamics of the proton Weibel instability will be the focus of a future study.

Finally, we end with a note that the results of this study prompt an additional inquiry analogous to the original studies of quasi-thermal noise in PIC codes, such as the numerical fluctuation-dissipation relation derived in \citet{Langdon:1979}.
All of the previous work in quantifying noise in the PIC method has focused on the uncorrelated charge density fluctuations and the corresponding noise-driven electrostatic field from these fluctuations.
While uncorrelated charge density fluctuations lead to uncorrelated current density fluctuations via the equation of charge continuity, and violations of charge continuity are themselves a source of noise which can be mitigated via additional improvements to the particle-in-cell method such as energy-conserving methods \citep{Markidis:2011}, we note that the \pic~simulations considered here employ Marder's correction \citep{Marder:1987} to the charge density to mitigate such errors.
Thus, we take as motivation this observed noise-driven magnetic field to propose an extension of the previous work quantifying noise in the PIC method, rederiving the modern results for the continuum magnetic field spectra of real plasmas \citep{Yoon:2007, Schlickeiser:2012, Yoon:2014} but using particles with finite shape.
In much the same way as the results of \citet{Langdon:1979} permit PIC codes to carefully filter and control electrostatic noise \citep{Haggerty:2017} by providing precise predictions of the noise generated by density fluctuations of the macroparticles, a similar prediction could be calculated from a fluctuation-dissipation relation on the magnitude of magnetic field fluctuations generated by current density fluctuations of the macroparticles.

\section{Acknowledgements}
The authors wish to thank G. W. Hammett and W. Dorland for enlightening conversations on estimating particle noise, as well as the insights of the entire \gke~team.
This work used the Extreme Science and Engineering Discovery Environment (XSEDE), which is supported by National Science Foundation grant number ACI-1548562, and resources of the National Energy Research Scientific Computing Center (NERSC), a U.S. Department of Energy Office of Science User Facility operated under Contract No. DE-AC02-05CH11231..
J. Juno was supported by a NASA Earth and Space Science Fellowship (Grant No. 80NSSC17K0428).
M. Swisdak was supported by NASA Grant No. 80NSSC19K039.
J. M. TenBarge was supported by NSF SHINE award AGS-1622306.
V. Skoutnev is supported by the NSF Grant for the Max Planck Princeton Center (MPPC).
A. Hakim is supported by the High-Fidelity Boundary Plasma Simulation SciDAC Project, part of the DOE Scientific Discovery Through Advanced Computing (SciDAC) program, through the U.S. Department of Energy contract No. DE-AC02-09CH11466 for the Princeton Plasma Physics Laboratory, and by Air Force Office of Scientific Research under Grant No. FA9550-15-1-0193.

\bibliographystyle{jpp}
\bibliography{abbrev.bib,picvlasov.bib}

\begin{thebibliography}{75}
\expandafter\ifx\csname natexlab\endcsname\relax\def\natexlab#1{#1}\fi
\def\au#1{#1} \def\ed#1{#1} \def\yr#1{#1}\def\at#1{#1}\def\jt#1{\textit{#1}}
  \def\bt#1{#1}\def\bvol#1{\textbf{#1}} \def\vol#1{#1} \def\pg#1{#1}
  \def\publ#1{#1}\def\arxiv#1{#1}\def\org#1{#1}\def\st#1{\textit{#1}}

\bibitem[Abel {\em et~al.\/}(2012)Abel, Hahn \& Kaehler]{Abel:2012}
{\sc \au{Abel, Tom}, \au{Hahn, Oliver} \& \au{Kaehler, Ralf}} \yr{2012}
  \at{{Tracing the dark matter sheet in phase space}}.
  \jt{Mon.~Not.~Roy.~Astron.~Soc.}  \bvol{427}~(1),  \pg{61--76}.

\bibitem[Arnold \& Awanou(2011)]{Arnold:2011}
{\sc \au{Arnold, Douglas~N} \& \au{Awanou, Gerard}} \yr{2011}  \at{{The
  Serendipity Family of Finite Elements}}.  \jt{Found.~Comput.~Math.}
  \bvol{11}~(3),  \pg{337--344}.

\bibitem[Barnes {\em et~al.\/}(2010)Barnes, Dorland \& Tatsuno]{Barnes:2010}
{\sc \au{Barnes, M.}, \au{Dorland, W.} \& \au{Tatsuno, T.}} \yr{2010}
  \at{Resolving velocity space dynamics in continuum gyrokinetics}.
  \jt{Phys.~Plasmas}  \bvol{17}~(3),  \pg{032106}.

\bibitem[Belova {\em et~al.\/}(1997)Belova, Denton \& Chan]{Belova:1997}
{\sc \au{Belova, E.V.}, \au{Denton, R.E.} \& \au{Chan, A.A.}} \yr{1997}
  \at{Hybrid simulations of the effects of energetic particles on low-frequency
  mhd waves}.  \jt{J.~Comp.~Phys.}  \bvol{136}~(2),  \pg{324--336}.

\bibitem[{Biermann}(1950)]{Biermann:1950}
{\sc \au{{Biermann}, L.}} \yr{1950}  \at{{{\"U}ber den Ursprung der
  Magnetfelder auf Sternen und im interstellaren Raum (miteinem Anhang von A.
  Schl{\"u}ter)}}.  \jt{Z.~Naturforsch.~Teil~A}  \bvol{5},  \pg{65}.

\bibitem[Birdsall \& Langdon(1990)]{birdsallbook}
{\sc \au{Birdsall, C.K} \& \au{Langdon, A.~B}} \yr{1990} {\em {Plasma Physics
  Via Computer Simulation}\/}.  \publ{Institute of Physics Publishing}.

\bibitem[Bornatici \& Lee(1970)]{Bornatici:1970}
{\sc \au{Bornatici, M.} \& \au{Lee, Kai~Fong}} \yr{1970}  \at{Ordinary-mode
  electromagnetic instability in counterstreaming plasmas with anisotropic
  temperatures}.  \jt{Phys.~Fluids}  \bvol{13}~(12),  \pg{3007--3011}.

\bibitem[Bowers {\em et~al.\/}(2009)Bowers, Albright, Yin, Daughton,
  Roytershteyn, Bergen \& Kwan]{Bowers:2009}
{\sc \au{Bowers, K~J}, \au{Albright, B~J}, \au{Yin, L}, \au{Daughton, W},
  \au{Roytershteyn, V}, \au{Bergen, B} \& \au{Kwan, T J~T}} \yr{2009}
  \at{Advances in petascale kinetic plasma simulation with {VPIC} and
  roadrunner}.  \jt{Journal of Physics: Conference Series}  \bvol{180},
  \pg{012055}.

\bibitem[Bret(2009)]{Bret:2009}
{\sc \au{Bret, A.}} \yr{2009}  \at{Weibel, two--stream, filamentation, oblique,
  bell, buneman... which one grows faster?}  \jt{Astrophys.~J.}
  \bvol{699}~(2),  \pg{990--1003}.

\bibitem[Cagas {\em et~al.\/}(2017)Cagas, Hakim, Scales \&
  Srinivasan]{Cagas:2017}
{\sc \au{Cagas, P.}, \au{Hakim, A.}, \au{Scales, W.} \& \au{Srinivasan, B.}}
  \yr{2017}  \at{Nonlinear saturation of the weibel instability}.
  \jt{Phys.~Plasmas}  \bvol{24}~(11),  \pg{112116}.

\bibitem[Califano {\em et~al.\/}(1997)Califano, Pegoraro \&
  Bulanov]{Califano:1997}
{\sc \au{Califano, F.}, \au{Pegoraro, F.} \& \au{Bulanov, S.~V.}} \yr{1997}
  \at{Spatial structure and time evolution of the weibel instability in
  collisionless inhomogeneous plasmas}.  \jt{Phys.~Rev.~E}  \bvol{56},
  \pg{963--969}.

\bibitem[Califano {\em et~al.\/}(1998)Califano, Prandi, Pegoraro \&
  Bulanov]{Califano:1998}
{\sc \au{Califano, F.}, \au{Prandi, R.}, \au{Pegoraro, F.} \& \au{Bulanov,
  S.~V.}} \yr{1998}  \at{Nonlinear filamentation instability driven by an
  inhomogeneous current in a collisionless plasma}.  \jt{Phys.~Rev.~E}
  \bvol{58},  \pg{7837--7845}.

\bibitem[Camporeale {\em et~al.\/}(2016)Camporeale, Delzanno, Bergen \&
  Moulton]{Camporeale:2016}
{\sc \au{Camporeale, E.}, \au{Delzanno, G.~L.}, \au{Bergen, B.~K.} \&
  \au{Moulton, J.~D.}} \yr{2016}  \at{{On the velocity space discretization for
  the Vlasov-Poisson system: Comparison between implicit Hermite spectral and
  Particle-in-Cell methods}}.  \jt{Comp.~Phys.~Comm.}  \bvol{198},
  \pg{47--58}.

\bibitem[Cheng {\em et~al.\/}(2013)Cheng, Parker, Chen \&
  Uzdensky]{ChengJianhua:2013}
{\sc \au{Cheng, Jianhua}, \au{Parker, Scott~E.}, \au{Chen, Yang} \&
  \au{Uzdensky, Dmitri~A.}} \yr{2013}  \at{A second-order semi-implicit $\delta
  f$ method for hybrid simulation}.  \jt{J.~Comp.~Phys.}  \bvol{245},
  \pg{364--375}.

\bibitem[Coppa {\em et~al.\/}(1996)Coppa, Lapenta, Dellapiana, Donato \&
  Riccardo]{Coppa:1996}
{\sc \au{Coppa, G.G.M.}, \au{Lapenta, G.}, \au{Dellapiana, G.}, \au{Donato, F.}
  \& \au{Riccardo, V.}} \yr{1996}  \at{Blob method for kinetic plasma
  simulation with variable-size particles}.  \jt{J.~Comp.~Phys.}
  \bvol{127}~(2),  \pg{268--284}.

\bibitem[Davidson {\em et~al.\/}(1972)Davidson, Hammer, Haber \&
  Wagner]{Davidson:1972}
{\sc \au{Davidson, Ronald~C.}, \au{Hammer, David~A.}, \au{Haber, Irving} \&
  \au{Wagner, Carl~E.}} \yr{1972}  \at{Nonlinear development of electromagnetic
  instabilities in anisotropic plasmas}.  \jt{Phys.~Fluids}  \bvol{15}~(2),
  \pg{317--333}.

\bibitem[Dawson(1962)]{Dawson:1962}
{\sc \au{Dawson, John}} \yr{1962}  \at{One-dimensional plasma model}.
  \jt{Phys.~Fluids}  \bvol{5}~(4),  \pg{445--459}.

\bibitem[Dawson(1983)]{Dawson:1983}
{\sc \au{Dawson, John~M.}} \yr{1983}  \at{Particle simulation of plasmas}.
  \jt{Rev. Mod. Phys.}  \bvol{55},  \pg{403--447}.

\bibitem[Denton \& Kotschenreuther(1995)]{Denton:1995}
{\sc \au{Denton, Richard~E.} \& \au{Kotschenreuther, M.}} \yr{1995}
  \at{$\delta f$ algorithm}.  \jt{J.~Comp.~Phys.}  \bvol{119}~(2),
  \pg{283--294}.

\bibitem[Fiuza {\em et~al.\/}(2012)Fiuza, Fonseca, Tonge, Mori \&
  Silva]{Fiuza:2012}
{\sc \au{Fiuza, F.}, \au{Fonseca, R.~A.}, \au{Tonge, J.}, \au{Mori, W.~B.} \&
  \au{Silva, L.~O.}} \yr{2012}  \at{Weibel-instability-mediated collisionless
  shocks in the laboratory with ultraintense lasers}.  \jt{Phys.~Rev.~Lett.}
  \bvol{108},  \pg{235004}.

\bibitem[Fonseca {\em et~al.\/}(2008)Fonseca, Martins, Silva, Tonge, Tsung \&
  Mori]{Fonseca:2008}
{\sc \au{Fonseca, R~A}, \au{Martins, S~F}, \au{Silva, L~O}, \au{Tonge, J~W},
  \au{Tsung, F~S} \& \au{Mori, W~B}} \yr{2008}  \at{One-to-one direct modeling
  of experiments and astrophysical scenarios: pushing the envelope on kinetic
  plasma simulations}.  \jt{Plasma Phys.~Con.~Fus.}  \bvol{50}~(12),
  \pg{124034}.

\bibitem[Fonseca {\em et~al.\/}(2003)Fonseca, Silva, Tonge, Mori \&
  Dawson]{Fonseca:2003}
{\sc \au{Fonseca, Ricardo~A.}, \au{Silva, Lu\'is~O.}, \au{Tonge, John~W.},
  \au{Mori, Warren~B.} \& \au{Dawson, John~M.}} \yr{2003}
  \at{Three-dimensional weibel instability in astrophysical scenarios}.
  \jt{Phys.~Plasmas}  \bvol{10}~(5),  \pg{1979--1984}.

\bibitem[Fried(1959)]{Fried:1959}
{\sc \au{Fried, Burton~D.}} \yr{1959}  \at{Mechanism for instability of
  transverse plasma waves}.  \jt{Phys.~Fluids}  \bvol{2}~(3),  \pg{337--337}.

\bibitem[Germaschewski {\em et~al.\/}(2016)Germaschewski, Fox, Abbott, Ahmadi,
  Maynard, Wang, Ruhl \& Bhattacharjee]{Germaschewski:2016}
{\sc \au{Germaschewski, Kai}, \au{Fox, William}, \au{Abbott, Stephen},
  \au{Ahmadi, Narges}, \au{Maynard, Kristofor}, \au{Wang, Liang}, \au{Ruhl,
  Hartmut} \& \au{Bhattacharjee, Amitava}} \yr{2016}  \at{The plasma simulation
  code: A modern particle-in-cell code with patch-based load-balancing}.
  \jt{J.~Comp.~Phys.}  \bvol{318},  \pg{305--326}.

\bibitem[Haggerty {\em et~al.\/}(2017)Haggerty, Parashar, Matthaeus, Shay,
  Yang, Wan, Wu \& Servidio]{Haggerty:2017}
{\sc \au{Haggerty, C.~C.}, \au{Parashar, T.~N.}, \au{Matthaeus, W.~H.},
  \au{Shay, M.~A.}, \au{Yang, Y.}, \au{Wan, M.}, \au{Wu, P.} \& \au{Servidio,
  S.}} \yr{2017}  \at{Exploring the statistics of magnetic reconnection
  x-points in kinetic particle-in-cell turbulence}.  \jt{Phys.~Plasmas}
  \bvol{24}~(10),  \pg{102308}.

\bibitem[Hahn \& Angulo(2015)]{Hahn:2015}
{\sc \au{Hahn, Oliver} \& \au{Angulo, Raul~E.}} \yr{2015}  \at{{An adaptively
  refined phase–space element method for cosmological simulations and
  collisionless dynamics}}.  \jt{Mon.~Not.~Roy.~Astron.~Soc.}  \bvol{455}~(1),
  \pg{1115--1133}.

\bibitem[Hakim {\em et~al.\/}(2019)Hakim, Francisquez, Juno \&
  Hammett]{Hakim:2019}
{\sc \au{Hakim, Ammar}, \au{Francisquez, M.}, \au{Juno, J.} \& \au{Hammett,
  Greg~W.}} \yr{2019}  \at{Conservative discontinuous galerkin schemes for
  nonlinear fokker-planck collision operators} ,  \arxiv{arXiv:
  arXiv:1903.08062}.

\bibitem[Hockney(1971)]{Hockney:1971}
{\sc \au{Hockney, R.W}} \yr{1971}  \at{Measurements of collision and heating
  times in a two-dimensional thermal computer plasma}.  \jt{J.~Comp.~Phys.}
  \bvol{8}~(1),  \pg{19--44}.

\bibitem[Hockney(1968)]{Hockney:1968}
{\sc \au{Hockney, R.~W.}} \yr{1968}  \at{Characteristics of noise in a
  two-dimensional computer plasma}.  \jt{Phys.~Fluids}  \bvol{11}~(6),
  \pg{1381--1383}.

\bibitem[Hu \& Krommes(1994)]{Hu:1994}
{\sc \au{Hu, Genze} \& \au{Krommes, John~A.}} \yr{1994}  \at{Generalized
  weighting scheme for $\delta f$ particle simulation method}.
  \jt{Phys.~Plasmas}  \bvol{1}~(4),  \pg{863--874}.

\bibitem[Huntington {\em et~al.\/}(2015)Huntington, Fiuza, Ross, Zylstra,
  Drake, Froula, Gregori, Kugland, Kuranz, Levy, Li, Meinecke, Morita,
  Petrasso, Plechaty, Remington, Ryutov, Sakawa, Spitkovsky, Takabe \&
  Park]{Huntington:2015}
{\sc \au{Huntington, C.~M.}, \au{Fiuza, F.}, \au{Ross, J.~S.}, \au{Zylstra,
  A.~B.}, \au{Drake, R.~P.}, \au{Froula, D.~H.}, \au{Gregori, G.}, \au{Kugland,
  N.~L.}, \au{Kuranz, C.~C.}, \au{Levy, M.~C.}, \au{Li, C.~K.}, \au{Meinecke,
  J.}, \au{Morita, T.}, \au{Petrasso, R.}, \au{Plechaty, C.}, \au{Remington,
  B.~A.}, \au{Ryutov, D.~D.}, \au{Sakawa, Y.}, \au{Spitkovsky, A.}, \au{Takabe,
  H.} \& \au{Park, H.~S.}} \yr{2015}  \at{Observation of magnetic field
  generation via the weibel instability in interpenetrating plasma flows}.
  \jt{Nature}  \bvol{11}~(2),  \pg{173--176}.

\bibitem[{Juno} {\em et~al.\/}(2018){Juno}, {Hakim}, {TenBarge}, {Shi} \&
  {Dorland}]{Juno:2018}
{\sc \au{{Juno}, J.}, \au{{Hakim}, A.}, \au{{TenBarge}, J.}, \au{{Shi}, E.} \&
  \au{{Dorland}, W.}} \yr{2018}  \at{{Discontinuous Galerkin algorithms for
  fully kinetic plasmas}}.  \jt{J.~Comp.~Phys.}  \bvol{353},  \pg{110--147}.

\bibitem[Kates-Harbeck {\em et~al.\/}(2016)Kates-Harbeck, Totorica, Zrake \&
  Abel]{KatesHarbeck:2016}
{\sc \au{Kates-Harbeck, Julian}, \au{Totorica, Samuel}, \au{Zrake, Jonathan} \&
  \au{Abel, Tom}} \yr{2016}  \at{Simplex-in-cell technique for collisionless
  plasma simulations}.  \jt{J.~Comp.~Phys.}  \bvol{304},  \pg{231--251}.

\bibitem[Kato \& Takabe(2008)]{Kato:2008}
{\sc \au{Kato, Tsunehiko~N.} \& \au{Takabe, Hideaki}} \yr{2008}
  \at{Nonrelativistic collisionless shocks in unmagnetized electron-ion
  plasmas}.  \jt{Astrophys.~J.~Lett.}  \bvol{681}~(2),  \pg{L93--L96}.

\bibitem[Kazimura {\em et~al.\/}(1998)Kazimura, Sakai, Neubert \&
  Bulanov]{Kazimura:1998}
{\sc \au{Kazimura, Y.}, \au{Sakai, J.~I.}, \au{Neubert, T.} \& \au{Bulanov,
  S.~V.}} \yr{1998}  \at{Generation of a small-scale quasi-static magnetic
  field and fast particles during the collision of electron-positron plasma
  clouds}.  \jt{Astrophys.~J.~Lett.}  \bvol{498}~(2),  \pg{L183--L186}.

\bibitem[Krommes(2007)]{Krommes:2007}
{\sc \au{Krommes, John~A.}} \yr{2007}  \at{Nonequilibrium gyrokinetic
  fluctuation theory and sampling noise in gyrokinetic particle-in-cell
  simulations}.  \jt{Phys.~Plasmas}  \bvol{14}~(9),  \pg{090501}.

\bibitem[Kumar {\em et~al.\/}(2015)Kumar, Eichler \& Gedalin]{Kumar:2015}
{\sc \au{Kumar, Rahul}, \au{Eichler, David} \& \au{Gedalin, Michael}} \yr{2015}
   \at{Electron heating in a relativistic, weibel-unstable plasma}.
  \jt{Astrophys.~J.}  \bvol{806}~(2),  \pg{165}.

\bibitem[Kunz {\em et~al.\/}(2014)Kunz, Stone \& Bai]{Kunz:2014b}
{\sc \au{Kunz, Matthew~W.}, \au{Stone, James~M.} \& \au{Bai, Xue-Ning}}
  \yr{2014}  \at{{Pegasus: A new hybrid-kinetic particle-in-cell code for
  astrophysical plasma dynamics}}.  \jt{J.~Comp.~Phys.}  \bvol{259},
  \pg{154--174}.

\bibitem[Langdon(1979)]{Langdon:1979}
{\sc \au{Langdon, A.~Bruce}} \yr{1979}  \at{Kinetic theory for fluctuations and
  noise in computer simulation of plasma}.  \jt{Phys.~Fluids}  \bvol{22}~(1),
  \pg{163--171}.

\bibitem[Langdon \& Birdsall(1970)]{Langdon:1970}
{\sc \au{Langdon, A.~Bruce} \& \au{Birdsall, Charles~K.}} \yr{1970}  \at{Theory
  of plasma simulation using finite-size particles}.  \jt{Phys.~Fluids}
  \bvol{13}~(8),  \pg{2115--2122}.

\bibitem[Lapenta(2012)]{Lapenta:2012}
{\sc \au{Lapenta, Giovanni}} \yr{2012}  \at{{Particle simulations of space
  weather}}.  \jt{J.~Comp.~Phys.}  \bvol{231}~(3),  \pg{795--821}.

\bibitem[Lazar {\em et~al.\/}(2009)Lazar, Schlickeiser, Wielebinski \&
  Poedts]{Lazar:2009}
{\sc \au{Lazar, M.}, \au{Schlickeiser, R.}, \au{Wielebinski, R.} \& \au{Poedts,
  S.}} \yr{2009}  \at{Cosmological effects of weibel-type instabilities}.
  \jt{Astrophys.~J.}  \bvol{693}~(2),  \pg{1133--1141}.

\bibitem[Marder(1987)]{Marder:1987}
{\sc \au{Marder, Barry}} \yr{1987}  \at{A method for incorporating gauss' law
  into electromagnetic pic codes}.  \jt{J.~Comp.~Phys.}  \bvol{68}~(1),
  \pg{48--55}.

\bibitem[Markidis \& Lapenta(2011)]{Markidis:2011}
{\sc \au{Markidis, Stefano} \& \au{Lapenta, Giovanni}} \yr{2011}  \at{The
  energy conserving particle-in-cell method}.  \jt{J.~Comp.~Phys.}
  \bvol{230}~(18),  \pg{7037--7052}.

\bibitem[Matteucci {\em et~al.\/}(2019)Matteucci, Fox, Bhattacharjee,
  Schaeffer, Germaschewski, Fiksel, Peery \& Hu]{Matteucci:2019}
{\sc \au{Matteucci, J.}, \au{Fox, W.}, \au{Bhattacharjee, A.}, \au{Schaeffer,
  D.~B.}, \au{Germaschewski, K.}, \au{Fiksel, G.}, \au{Peery, J.} \& \au{Hu,
  S.~X.}} \yr{2019}  \at{{Weibel instability drives large magnetic field
  generation in laser-driven single plume ablation}} ,  \arxiv{arXiv:
  1912.11120}.

\bibitem[Medvedev \& Loeb(1999)]{Medvedev:1999}
{\sc \au{Medvedev, Mikhail~V.} \& \au{Loeb, Abraham}} \yr{1999}  \at{Generation
  of magnetic fields in the relativistic shock of gamma-ray burst sources}.
  \jt{Astrophys.~J.}  \bvol{526}~(2),  \pg{697--706}.

\bibitem[Meyer-Vernet {\em et~al.\/}(1986)Meyer-Vernet, Couturier, Hoang,
  Perche, Steinberg, Fainberg \& Meetre]{MeyerVernet:1986}
{\sc \au{Meyer-Vernet, N.}, \au{Couturier, P.}, \au{Hoang, S.}, \au{Perche,
  C.}, \au{Steinberg, J.~L.}, \au{Fainberg, J.} \& \au{Meetre, C.}} \yr{1986}
  \at{Plasma diagnosis from thermal noise and limits on dust flux or mass in
  comet giacobini-zinner}.  \jt{Sci.}  \bvol{232}~(4748),  \pg{370--374}.

\bibitem[Morse \& Nielson(1971)]{Morse:1971}
{\sc \au{Morse, R.~L.} \& \au{Nielson, C.~W.}} \yr{1971}  \at{Numerical
  simulation of the weibel instability in one and two dimensions}.
  \jt{Phys.~Fluids}  \bvol{14}~(4),  \pg{830--840}.

\bibitem[Nevins {\em et~al.\/}(2005)Nevins, Hammett, Dimits, Dorland \&
  Shumaker]{Nevins:2005}
{\sc \au{Nevins, W.~M.}, \au{Hammett, G.~W.}, \au{Dimits, A.~M.}, \au{Dorland,
  W.} \& \au{Shumaker, D.~E.}} \yr{2005}  \at{Discrete particle noise in
  particle-in-cell simulations of plasma microturbulence}.  \jt{Phys.~Plasmas}
  \bvol{12}~(12),  \pg{122305}.

\bibitem[Nishikawa {\em et~al.\/}(2003)Nishikawa, Hardee, Richardson, Preece,
  Sol \& Fishman]{Nishikawa:2003}
{\sc \au{Nishikawa, K.-I.}, \au{Hardee, P.}, \au{Richardson, G.}, \au{Preece,
  R.}, \au{Sol, H.} \& \au{Fishman, G.~J.}} \yr{2003}  \at{Particle
  acceleration in relativistic jets due to weibel instability}.
  \jt{Astrophys.~J.}  \bvol{595}~(1),  \pg{555--563}.

\bibitem[Nishikawa {\em et~al.\/}(2005)Nishikawa, Hardee, Richardson, Preece,
  Sol \& Fishman]{Nishikawa:2005}
{\sc \au{Nishikawa, K.-I.}, \au{Hardee, P.}, \au{Richardson, G.}, \au{Preece,
  R.}, \au{Sol, H.} \& \au{Fishman, G.~J.}} \yr{2005}  \at{Particle
  acceleration and magnetic field generation in electron-positron relativistic
  shocks}.  \jt{Astrophys.~J.}  \bvol{622}~(2),  \pg{927--937}.

\bibitem[Okuda(1972)]{Okuda:1972}
{\sc \au{Okuda, Hideo}} \yr{1972}  \at{Verification of theory for plasma of
  finite-size particles}.  \jt{Phys.~Fluids}  \bvol{15}~(7),  \pg{1268--1274}.

\bibitem[Okuda \& Birdsall(1970)]{Okuda:1970}
{\sc \au{Okuda, Hideo} \& \au{Birdsall, Charles~K.}} \yr{1970}  \at{Collisions
  in a plasma of finite--size particles}.  \jt{Phys.~Fluids}  \bvol{13}~(8),
  \pg{2123--2134}.

\bibitem[Parker \& Lee(1993)]{Parker:1993}
{\sc \au{Parker, S.~E.} \& \au{Lee, W.~W.}} \yr{1993}  \at{A fully nonlinear
  characteristic method for gyrokinetic simulation}.  \jt{Phys.~Fluids B}
  \bvol{5}~(1),  \pg{77--86}.

\bibitem[Pusztai {\em et~al.\/}(2020)Pusztai, Juno, Brandenburg, TenBarge,
  Hakim, Francisquez \& Sundstr\"om]{Pusztai:2020}
{\sc \au{Pusztai, I.}, \au{Juno, J.}, \au{Brandenburg, A.}, \au{TenBarge,
  J.~M.}, \au{Hakim, A.~H.}, \au{Francisquez, M.} \& \au{Sundstr\"om, A.}}
  \yr{2020}  \at{{Dynamo in weakly collisional non-magnetized plasmas impeded
  by Landau damping of magnetic fields}} ,  \arxiv{arXiv: 2001.11929}.

\bibitem[Pusztai {\em et~al.\/}(2018)Pusztai, TenBarge, Csap\'o, Juno, Hakim,
  Yi \& F\"ul\"op]{Pusztai:2018}
{\sc \au{Pusztai, I.}, \au{TenBarge, J.~M.}, \au{Csap\'o, A.~N.}, \au{Juno,
  J.}, \au{Hakim, A.}, \au{Yi, L.} \& \au{F\"ul\"op, T.}} \yr{2018}  \at{Low
  mach-number collisionless electrostatic shocks and associated ion
  acceleration}.  \jt{Plasma Phys.~Con.~Fus.}  \bvol{60}~(3),  \pg{035004}.

\bibitem[Schlickeiser \& Shukla(2003)]{Schlickeiser:2003}
{\sc \au{Schlickeiser, R.} \& \au{Shukla, P.~K.}} \yr{2003}  \at{Cosmological
  magnetic field generation by the weibel instability}.
  \jt{Astrophys.~J.~Lett.}  \bvol{599}~(2),  \pg{L57--L60}.

\bibitem[Schlickeiser \& Yoon(2012)]{Schlickeiser:2012}
{\sc \au{Schlickeiser, R.} \& \au{Yoon, P.~H.}} \yr{2012}  \at{Spontaneous
  electromagnetic fluctuations in unmagnetized plasmas i: General theory and
  nonrelativistic limit}.  \jt{Phys.~Plasmas}  \bvol{19}~(2),  \pg{022105}.

\bibitem[Schoeffler {\em et~al.\/}(2014)Schoeffler, Loureiro, Fonseca \&
  Silva]{Schoeffler:2014}
{\sc \au{Schoeffler, K.~M.}, \au{Loureiro, N.~F.}, \au{Fonseca, R.~A.} \&
  \au{Silva, L.~O.}} \yr{2014}  \at{Magnetic-field generation and amplification
  in an expanding plasma}.  \jt{Phys.~Rev.~Lett.}  \bvol{112},  \pg{175001}.

\bibitem[Schoeffler {\em et~al.\/}(2016)Schoeffler, Loureiro, Fonseca \&
  Silva]{Schoeffler:2016}
{\sc \au{Schoeffler, K.~M.}, \au{Loureiro, N.~F.}, \au{Fonseca, R.~A.} \&
  \au{Silva, L.~O.}} \yr{2016}  \at{The generation of magnetic fields by the
  biermann battery and the interplay with the weibel instability}.
  \jt{Phys.~Plasmas}  \bvol{23}~(5),  \pg{056304}.

\bibitem[Schoeffler {\em et~al.\/}(2018)Schoeffler, Loureiro \&
  Silva]{Schoeffler:2018}
{\sc \au{Schoeffler, K.~M.}, \au{Loureiro, N.~F.} \& \au{Silva, L.~O.}}
  \yr{2018}  \at{Fully kinetic biermann battery and associated generation of
  pressure anisotropy}.  \jt{Phys.~Rev.~E}  \bvol{97},  \pg{033204}.

\bibitem[Silva {\em et~al.\/}(2003)Silva, Fonseca, Tonge, Dawson, Mori \&
  Medvedev]{Silva:2003}
{\sc \au{Silva, L.~O.}, \au{Fonseca, R.~A.}, \au{Tonge, J.~W.}, \au{Dawson,
  J.~M.}, \au{Mori, W.~B.} \& \au{Medvedev, M.~V.}} \yr{2003}
  \at{Interpenetrating plasma shells: Near-equipartition magnetic field
  generation and nonthermal particle acceleration}.  \jt{Astrophys.~J.~Lett.}
  \bvol{596}~(1),  \pg{L121--L124}.

\bibitem[Skoutnev {\em et~al.\/}(2019)Skoutnev, Hakim, Juno \&
  TenBarge]{Skoutnev:2019}
{\sc \au{Skoutnev, V.}, \au{Hakim, A.}, \au{Juno, J.} \& \au{TenBarge, J.~M.}}
  \yr{2019}  \at{Temperature-dependent saturation of weibel-type instabilities
  in counter-streaming plasmas}.  \jt{Astrophys.~J.~Lett.}  \bvol{872}~(2),
  \pg{L28}.

\bibitem[{Spitkovsky}(2005)]{Spitkovsky:2005}
{\sc \au{{Spitkovsky}, Anatoly}} \yr{2005} {Simulations of relativistic
  collisionless shocks: shock structure and particle acceleration}.  \bt{In
  {\em Astrophysical Sources of High Energy Particles and Radiation\/} (ed.
  \ed{Tomasz {Bulik}, Bronislaw {Rudak} \& Grzegorz {Madejski}})},
  \st{American Institute of Physics Conference Series},  \vol{vol. 801},
  \pg{pp. 345--350}.

\bibitem[Sundstr\"om {\em et~al.\/}(2019)Sundstr\"om, Juno, TenBarge \&
  Pusztai]{Sundstrom:2019}
{\sc \au{Sundstr\"om, A.}, \au{Juno, J.}, \au{TenBarge, J.~M.} \& \au{Pusztai,
  I.}} \yr{2019}  \at{Effect of a weak ion collisionality on the dynamics of
  kinetic electrostatic shocks}.  \jt{J.~Plasma Phys.}  \bvol{85}~(1),
  \pg{905850108}.

\bibitem[{Tajima} {\em et~al.\/}(1992){Tajima}, {Cable}, {Shibata} \&
  {Kulsrud}]{Tajima:1992}
{\sc \au{{Tajima}, T.}, \au{{Cable}, S.}, \au{{Shibata}, K.} \& \au{{Kulsrud},
  R.~M.}} \yr{1992}  \at{{On the Origin of Cosmological Magnetic Fields}}.
  \jt{Astrophys.~J.}  \bvol{390},  \pg{309}.

\bibitem[Takamoto {\em et~al.\/}(2018)Takamoto, Matsumoto \&
  Kato]{Takamoto:2018}
{\sc \au{Takamoto, Makoto}, \au{Matsumoto, Yosuke} \& \au{Kato, Tsunehiko~N.}}
  \yr{2018}  \at{Magnetic field saturation of the ion weibel instability in
  interpenetrating relativistic plasmas}.  \jt{Astrophys.~J.~Lett.}
  \bvol{860}~(1),  \pg{L1}.

\bibitem[Totorica {\em et~al.\/}(2018)Totorica, Fiuza \& Abel]{Totorica:2018}
{\sc \au{Totorica, Samuel~R.}, \au{Fiuza, Frederico} \& \au{Abel, Tom}}
  \yr{2018}  \at{A new method for analyzing and visualizing plasma simulations
  using a phase-space tessellation}.  \jt{Phys.~Plasmas}  \bvol{25}~(7),
  \pg{072109}.

\bibitem[Weibel(1959)]{Weibel:1959}
{\sc \au{Weibel, Erich~S.}} \yr{1959}  \at{Spontaneously growing transverse
  waves in a plasma due to an anisotropic velocity distribution}.
  \jt{Phys.~Rev.~Lett.}  \bvol{2},  \pg{83--84}.

\bibitem[Wu \& Qin(2018)]{Wu:2018}
{\sc \au{Wu, Wentao} \& \au{Qin, Hong}} \yr{2018}  \at{Reducing noise for pic
  simulations using kernel density estimation algorithm}.  \jt{Phys.~Plasmas}
  \bvol{25}~(10),  \pg{102107}.

\bibitem[Nguyen~van yen {\em et~al.\/}(2010)Nguyen~van yen, del
  Castillo-Negrete, Schneider, Farge \& Chen]{vanyenNguyen:2010}
{\sc \au{Nguyen~van yen, Romain}, \au{del Castillo-Negrete, Diego},
  \au{Schneider, Kai}, \au{Farge, Marie} \& \au{Chen, Guangye}} \yr{2010}
  \at{Wavelet-based density estimation for noise reduction in plasma
  simulations using particles}.  \jt{J.~Comp.~Phys.}  \bvol{229}~(8),
  \pg{2821--2839}.

\bibitem[Nguyen~van yen {\em et~al.\/}(2011)Nguyen~van yen, Sonnendr\"ucker,
  Schneider \& Farge]{vanyenNguyen:2011}
{\sc \au{Nguyen~van yen, Romain}, \au{Sonnendr\"ucker, \'Eric}, \au{Schneider,
  Kai} \& \au{Farge, Marie}} \yr{2011}  \at{Particle-in-wavelets scheme for the
  1d vlasov-poisson equations}.  \jt{ESAIM: Proc.}  \bvol{32},  \pg{134--148}.

\bibitem[Yoon(2007)]{Yoon:2007}
{\sc \au{Yoon, Peter~H.}} \yr{2007}  \at{Spontaneous thermal magnetic field
  fluctuation}.  \jt{Phys.~Plasmas}  \bvol{14}~(6),  \pg{064504}.

\bibitem[Yoon {\em et~al.\/}(2014)Yoon, Schlickeiser \& Kolberg]{Yoon:2014}
{\sc \au{Yoon, P.~H.}, \au{Schlickeiser, R.} \& \au{Kolberg, U.}} \yr{2014}
  \at{Thermal fluctuation levels of magnetic and electric fields in
  unmagnetized plasma: The rigorous relativistic kinetic theory}.
  \jt{Phys.~Plasmas}  \bvol{21}~(3),  \pg{032109}.

\bibitem[Zeiler {\em et~al.\/}(2002)Zeiler, Biskamp, Drake, Rogers, Shay \&
  Scholer]{Zeiler:2002}
{\sc \au{Zeiler, A.}, \au{Biskamp, D.}, \au{Drake, J.~F.}, \au{Rogers, B.~N.},
  \au{Shay, M.~A.} \& \au{Scholer, M.}} \yr{2002}  \at{Three-dimensional
  particle simulations of collisionless magnetic reconnection}.
  \jt{J.~Geophys.~Res.}  \bvol{107}~(A9),  \pg{SMP 6--1--SMP 6--9}.

\end{thebibliography}

\end{document}